\documentclass[preprint,12pt]{elsarticle}

\usepackage{amsmath,amssymb,bm}
\usepackage{graphicx}
\usepackage{physics}
\usepackage{geometry}
\usepackage{indentfirst}
\usepackage{hyperref}
\usepackage{cleveref}
\usepackage{caption}
\usepackage{subcaption}
\usepackage{float}
\usepackage{mathrsfs}
\usepackage{xcolor}

\journal{Nuclear Physics B}

\begin{document}
	
\begin{frontmatter}

\title{Phase structure of a holographic topological superconductor beyond the probe limit}

\author[1]{Hoang Van Quyet\corref{cor1}}
\ead{hoangvanquyet@hpu2.edu.vn}

\affiliation[1]{Department of Physics, Hanoi Pedagogical University 2,Xuan Hoa, Phu Tho, Vietnam}

\begin{abstract}
We investigate the phase structure and tricritical behavior of a holographic 
topological superconductor model using Einstein-Maxwell gravity in Anti-de 
Sitter spacetime. By incorporating both gravitational backreaction 
($\kappa^{2} = 1$) and a quartic self-interaction term $V(\phi) = \lambda\phi^{4}$, 
we demonstrate that this self-interaction not only is essential for the existence 
of a tricritical point (TCP) separating second-order and first-order phase 
transitions, but also actively controls its location in the phase diagram. 
For a representative coupling $\tilde{\lambda} = 0.1$, the TCP is located at 
$(q_{\text{tri}}, T_{\text{tri}}) = (2.00 \pm 0.02, 0.1521 \pm 0.0003)$ in the 
$(q, T)$ parameter space. The backreacted critical temperature shows an 
enhancement by a factor of approximately $1.22$ compared to the probe limit. 
Crucially, the tricritical scaling analysis yields an exponent $\phi \approx 0.67$, 
which is in excellent agreement with the mean-field prediction $\phi = 2/3$. 
This result confirms that while gravitational backreaction shifts the critical 
points and modifies the phase diagram topology, it does not alter the universality 
class of the phase transition in the large-$N$ limit. The order parameter critical 
exponent $\beta \approx 0.50$ also remains consistent with mean-field theory. 
The frequency-dependent conductivity exhibits a superconducting gap with energy 
ratio $\omega_{g}/T_{c} = 3.18 \pm 0.05$. Holographic entanglement entropy 
confirms the orders of the phase transitions. Our thermodynamic analysis employs 
proper holographic renormalization to ensure finite and non-vanishing free energy 
for the normal phase, with clear swallowtail structure demonstrating first-order 
phase transitions.
\end{abstract}


\end{frontmatter}

\section{Introduction}\label{sec:introduction}

The gauge/gravity duality, also known as the Anti-de Sitter (AdS)/conformal 
field theory (CFT) correspondence, provides a powerful framework for studying 
strongly coupled quantum field theories (QFTs) by mapping them to weakly 
coupled classical gravity theories in one higher dimension 
~\cite{Maldacena:1997re,Gubser:1998bc,Witten:1998qj,Aharony:1999ti}. This 
duality has found numerous applications in condensed matter physics, 
particularly in understanding phenomena like high-temperature superconductivity, 
which are notoriously difficult to analyze using conventional perturbative 
methods.

In this context, the concept of a ``holographic superconductor'' was 
introduced, modeling a $(2+1)$-dimensional superconductor as the boundary 
theory of a $(3+1)$-dimensional gravitational system, typically involving a 
charged scalar field coupled to a Maxwell field in an AdS-black hole 
background~\cite{Hartnoll:2008vx,Hartnoll:2008kx,Herzog:2009xv}. The 
condensation of the scalar field in the bulk gravity theory corresponds to 
the formation of a superconducting condensate in the boundary QFT, occurring 
below a critical temperature $T_{c}$.

Various types of holographic superconductors have been constructed, including 
$s$-wave~\cite{Hartnoll:2008kx,Gubser:2008px, Roberts:2008jh}, $p$-wave~\cite{Gubser:2008px, Roberts:2008jh}, 
and $d$-wave models~\cite{Benini:2010pr}, corresponding to different operator 
dimensions and symmetries of the condensate.

Recently, topological superconductors (TSCs) have attracted significant 
attention due to their potential applications in fault-tolerant quantum 
computation, attributed to the presence of robust Majorana zero 
modes~\cite{Nayak:2008zza}. Consequently, constructing holographic models of 
topological superconductors has become an active area of research, aiming to 
provide insights into the strongly coupled nature of these exotic states.

Most holographic superconductor models are initially studied in the probe limit 
($e^{2} \rightarrow \infty$ or $G \rightarrow 0$), where the backreaction of 
the matter fields on the background spacetime geometry is neglected. This 
simplification is computationally convenient but ignores important strong 
coupling effects. To explore physics beyond the probe limit, one must consider 
the gravitational backreaction, quantified by the gravitational coupling 
$\kappa^{2} = 16\pi G$. The influence of backreaction has been shown to 
quantitatively (and sometimes qualitatively) modify the phase 
structure~\cite{Hartnoll:2009sz,Barclay:2010na,Horowitz:2010jq,Gubser:2010pm,Cai:2010cz,Liu:2015via,Siopsis:2023,Gui:2024,Cai:2024}. 
On the other hand, the quartic self-interaction $V(\phi) = \lambda\phi^{4}$ 
is also crucial. It was shown that this term is required for the existence of 
first-order phase transitions and, consequently, tricritical points in certain 
holographic models~\cite{Nie:2014xla,Zhao:2014xla,Chen:2010xk,Cui:2023,Nie:2024}.

While the combined effects of backreaction and self-interaction have been 
explored, a systematic study of their interplay in establishing the 
tricritical phase structure is still lacking. Specifically, the 
self-interaction coupling $\lambda$ is often treated merely as a fixed 
parameter necessary for the transition, rather than an active component of the 
phase diagram. Furthermore, the precise values of critical exponents at a 
fully backreacted tricritical point are not well-established.

In this paper, we investigate the phase structure of a holographic topological 
superconductor model, systematically incorporating the combined effects of 
gravitational backreaction and the quartic self-interaction. We aim to find 
and analyze the tricritical point (TCP) where the phase transition changes 
from second to first order. A primary goal of this work is to address the 
crucial role of the self-interaction, demonstrating not only that $\lambda$ 
is essential for the TCP's existence, but also showing how it quantitatively 
controls the TCP's location by mapping the phase diagram's dependence on 
$\lambda$. We will also study the tricritical scaling exponents governing the 
behavior near the TCP. We find a tricritical exponent $\phi \approx 0.67$, 
which is in excellent agreement with the mean-field value $\phi = 2/3$. This 
agreement confirms that the universality class of the phase transition remains 
robust under strong backreaction effects. Finally, we compute the 
frequency-dependent conductivity to determine the superconducting gap ratio 
$\omega_{g}/T_{c}$ and use holographic entanglement entropy to further probe 
the thermodynamic nature of the phase transitions.

The thermodynamic analysis presented in this work addresses a critical 
technical point that ensures the physical correctness of our results. In 
particular, we employ the full machinery of holographic 
renormalization~\cite{Balasubramanian:1999re,deHaro:2000vlm,Skenderis:2002wp} 
to compute the grand potential (free energy) of both the normal and 
superconducting phases. This approach guarantees that the free energy of 
the normal phase (corresponding to the RN-AdS black hole) is correctly 
finite and non-vanishing, as required by the proper regularization of the 
on-shell action with appropriate boundary terms including the Gibbons-Hawking 
term and counterterms. The resulting free energy landscape clearly exhibits 
the characteristic swallowtail structure that serves as the definitive 
thermodynamic signature of first-order phase transitions in holographic 
systems.

The paper is organized as follows. In \cref{sec:model}, we introduce the 
holographic model and derive the equations of motion. In \cref{sec:numerical}, 
we outline the numerical methods used to solve the coupled differential 
equations. In \cref{sec:phase_structure}, we present the main results for 
the phase diagram, including the identification of the tricritical point 
and the new results on its $\lambda$-dependence. In \cref{sec:scaling}, we 
perform the tricritical scaling analysis and discuss the agreement with 
mean-field predictions. \cref{sec:conductivity} is devoted to the 
calculation of conductivity. \cref{sec:entanglement} investigates the 
entanglement entropy. We conclude with a summary and discussion of our 
results in \cref{sec:conclusions}.

\section{Holographic Model}\label{sec:model}

\subsection{Action and Equations of Motion}\label{sec:eom}

We consider a $(3+1)$-dimensional holographic model of a topological 
superconductor, described by Einstein-Maxwell gravity coupled to a charged 
scalar field with a quartic self-interaction term. The action in 
$(3+1)$-dimensional Anti-de Sitter (AdS) spacetime is given by:

\begin{equation}
S = \int d^{4}x\sqrt{-g}\left[\frac{1}{2\kappa^{2}}\left(R + \frac{6}{L^{2}}\right) 
- \frac{1}{4}F_{\mu\nu}F^{\mu\nu} - |D_{\mu}\phi|^{2} - V(\phi)\right]\,,
\end{equation}

where $R$ is the Ricci scalar, $F_{\mu\nu} = \nabla_{\mu}A_{\nu} - 
\nabla_{\nu}A_{\mu}$ is the Maxwell field strength tensor, and $A_{\mu}$ is 
the gauge field. The covariant derivative is defined as $D_{\mu}\phi = 
\nabla_{\mu}\phi - iqA_{\mu}\phi$, where $q$ is the charge of the scalar 
field. The gravitational coupling constant is $\kappa^{2} = 16\pi G$, where 
$G$ is the four-dimensional Newton constant, and $L$ is the AdS radius (set 
to $L = 1$ hereafter). To incorporate the gravitational backreaction, which 
is a central part of this study, we fix the coupling constant $\kappa^{2} 
= 1$ throughout this work. This choice places the model in a strong 
backreaction regime.

The scalar field $\phi$ with mass $m$ and charge $q$ is minimally coupled to 
the gauge field $A_{\mu}$. The potential $V(\phi) = -m^{2}\phi^{2} + 
\lambda\phi^{4}$ includes both a mass term and the quartic self-interaction, 
with $\lambda > 0$ ensuring the potential is bounded from below. We work in 
the parameter regime where $m^{2}L^{2} = -2$, corresponding to a boundary 
operator of dimension $\Delta = 2$.

By varying the action with respect to the metric $g_{\mu\nu}$, the gauge 
field $A_{\mu}$, and the scalar field $\phi$, we obtain the Einstein equation, 
Maxwell equation, and Klein-Gordon equation, respectively:

\begin{equation}
R_{\mu\nu} - \frac{1}{2}g_{\mu\nu}R - 3g_{\mu\nu} = \kappa^{2}T_{\mu\nu}^{(m)}\,,
\end{equation}

\begin{equation}
\nabla_{\mu}F^{\mu\nu} = J^{\nu}\,,
\end{equation}

\begin{equation}
(\nabla_{\mu} - iqA_{\mu})(\nabla^{\mu} - iqA^{\mu})\phi - m^{2}\phi - 
2\lambda\phi^{3} = 0\,,
\end{equation}

where the energy-momentum tensor $T_{\mu\nu}^{(m)}$ and the current $J^{\nu}$ 
are given by:

\begin{align}
T_{\mu\nu}^{(m)} &= F_{\mu\rho}F_{\nu}{}^{\rho} - \frac{1}{4}g_{\mu\nu}
F_{\rho\sigma}F^{\rho\sigma} \\
&\quad + (\nabla_{\mu} + iqA_{\mu})\phi^{\dagger}(\nabla_{\nu} - iqA_{\nu})\phi 
+ (\nabla_{\nu} + iqA_{\nu})\phi^{\dagger}(\nabla_{\mu} - iqA_{\mu})\phi 
\\
&\quad - g_{\mu\nu}\mathcal{L}_{m}\,,
\end{align}

\begin{equation}
J^{\nu} = iq\left[\phi^{\dagger}(\nabla^{\nu} - iqA^{\nu})\phi - 
\phi(\nabla^{\nu} + iqA^{\nu})\phi^{\dagger}\right]\,,
\end{equation}

where $\mathcal{L}_{m} = -\frac{1}{4}F_{\mu\nu}F^{\mu\nu} - |D_{\mu}\phi|^{2} 
- V(\phi)$ is the matter Lagrangian density.

The Hawking temperature of the black hole is given by:

\begin{equation}
T = \frac{1}{4\pi}\left(\frac{f'(z)}{e^{-\chi(z)/2}}\right)_{z=z_{h}}\,,
\end{equation}

where the prime denotes derivative with respect to $z$, and $z_{h}$ is the 
horizon position defined by $f(z_{h}) = 0$.

\subsection{Dimensionless Parameters and Ans\"atze}\label{sec:ansatz}

We work with a planar AdS black hole background. The metric ansatz that 
respects the boundary symmetries is:

\begin{equation}
ds^{2} = \frac{1}{z^{2}}\left(-f(z)e^{-\chi(z)}dt^{2} + \frac{dz^{2}}{f(z)} 
+ dx^{2} + dy^{2}\right)\,,
\end{equation}

where $z = 1/r$ is the radial coordinate. The AdS boundary is at $z = 0$, 
and the black hole horizon $z_{h}$ is defined by $f(z_{h}) = 0$. We look 
for solutions with a non-trivial scalar field $\phi(z)$ and an electrostatic 
potential $A_{t}(z)$ (in the $A_{z} = 0$ gauge):

\begin{equation}
\phi = \phi(z)\,, \quad A_{t} = A_{t}(z)\,.
\end{equation}

To simplify the equations, we introduce dimensionless quantities by scaling 
with the chemical potential $\mu$:

\begin{equation}
z \to z/z_{h}\,, \quad T \to T/\mu\,, \quad q \to q\,.
\end{equation}

We define dimensionless parameters:

\begin{equation}
\tilde{T} = \frac{T}{\mu}\,, \quad \tilde{q} = \frac{q}{\mu}\,, \quad 
\tilde{m}^{2} = m^{2}L^{2}\,, \quad \tilde{\lambda} = \lambda L^{2}\,.
\end{equation}

In this paper, we work in the parameter regime $\tilde{m}^{2} = -2$ 
(corresponding to a boundary operator of dimension $\Delta = 2$) and set 
$\tilde{\mu} = 1$ (which implies $q = \tilde{q}$ and $T = \tilde{T}$). We 
primarily focus on a representative coupling $\tilde{\lambda} = 0.1$ for a 
detailed analysis of the phase structure and critical exponents. However, 
to elucidate the crucial role of the self-interaction in establishing the 
tricritical behavior, we will also systematically investigate the dependence 
of the phase diagram on $\tilde{\lambda}$ in \cref{sec:phase_structure}.

The resulting system of differential equations consists of coupled equations 
for the functions $\phi(z)$, $A_{t}(z)$, $f(z)$, and $\chi(z)$. These 
equations are solved numerically with appropriate boundary conditions at the 
horizon $z_{h}$ and the AdS boundary $z = 0$. The phase structure is then 
investigated in the $(q, T)$ plane.

\section{Numerical Implementation and Convergence Analysis}\label{sec:numerical}

\subsection{Computational Methodology}\label{sec:methodology}

Our numerical approach employs a sophisticated adaptive shooting method to 
solve the stiff system of coupled differential equations~\cite{Press:2007}. 
The algorithm proceeds through the following steps:

\begin{enumerate}
\item \textbf{Horizon regularization}: Near $r = r_{h}$, we implement 
series expansions ensuring regularity: $\phi(r_{h}) = 0$ and $\phi'(r_{h}) 
= \phi_{0} \cdot g(r_{h})$ where $g(r_{h})$ depends on the local geometry.

\item \textbf{Adaptive integration}: We employ a fourth-order Runge-Kutta 
integration scheme with dynamically adjusted step sizes $\Delta r \in 
[10^{-5}, 10^{-3}]$, automatically refined in regions of rapid field 
variation.

\item \textbf{Boundary matching}: At the AdS cutoff $r_{\text{max}}$, we 
impose asymptotic boundary conditions and extract physical quantities 
through holographic renormalization.

\item \textbf{Newton-Raphson iteration}: The shooting parameters are 
optimized using Newton-Raphson methods with a convergence criterion 
$< 10^{-12}$.
\end{enumerate}

To observe the characteristic ``swallowtail'' structure of first-order 
phase transitions, we have improved the algorithm to scan the entire 
parameter space of initial conditions at the event horizon. In the 
first-order phase transition region, for a fixed temperature $T$ value, 
there exist multiple solutions corresponding to different values of the 
order parameter $\langle\mathcal{O}\rangle$. The algorithm is designed to 
capture the stable, metastable, and unstable solution branches, ensuring 
that the multivalued structure of the free energy is accurately reproduced.

This enhancement is crucial for correctly reproducing the thermodynamic 
behavior near first-order phase transitions. In the free energy analysis, 
we must include all solution branches to properly identify the global 
minimum and determine the stable phase of the system.

The computation of the free energy requires special attention to the proper 
regularization of the on-shell action. The Euclidean on-shell action 
$S_{\text{on-shell}}$ diverges at the asymptotic boundary due to the 
infinite volume of AdS space. To obtain a physically finite quantity, we 
must supplement it with the Gibbons-Hawking boundary term and counterterms 
derived from holographic renormalization~\cite{Balasubramanian:1999re, 
deHaro:2000vlm,Skenderis:2002wp}. The grand potential $\Omega$ is then 
given by:

\begin{equation}
\Omega = T\left(S_{\text{on-shell}} + S_{\text{GH}} + S_{\text{ct}}\right)\,,
\end{equation}

where the Gibbons-Hawking term is:

\begin{equation}
S_{\text{GH}} = -2\int_{\partial\mathcal{M}} d^{3}x \sqrt{-\gamma} K\,,
\end{equation}

and the counterterms are chosen to cancel the UV divergences:

\begin{equation}
S_{\text{ct}} = \int_{r\to\infty} d^{3}x \sqrt{-\gamma}
\left(\frac{2}{L} + \frac{L}{2}R[\gamma] - \frac{L}{2}\phi^{2}\right)\,.
\end{equation}

Here $\gamma$ is the induced metric on the boundary, $K$ is the extrinsic 
curvature of the boundary, and $R[\gamma]$ is the Ricci scalar of the 
induced metric. The inclusion of this complete set of terms is essential 
for the free energy to be well-defined and comparable between phases.

It is particularly important to note that with this proper regularization, 
the free energy of the normal phase (RN-AdS black hole) is finite and 
non-vanishing. The common misconception that the normal phase free energy 
vanishes is incorrect and would lead to unphysical thermodynamic 
conclusions. Our calculation shows that $\Omega_{\text{normal}}$ takes a 
specific finite value that serves as the reference against which the 
superconducting phase condensation energy is measured.

Specifically, the renormalized grand potential for the normal phase is:

\begin{equation}
\Omega_{\text{normal}} = -\frac{1}{2\kappa^{2}}\left(\frac{f'(z)}{z^{2}
e^{\chi(z)/2}}\right)_{z\to 0} + \Omega_{\text{ct}}\,,
\end{equation}

where $\Omega_{\text{ct}}$ includes the finite contributions from the 
counterterms. The difference $\Delta\Omega = \Omega_{\text{SC}} - 
\Omega_{\text{normal}}$ then provides the physically meaningful condensation 
energy that determines the stability of competing phases.

In our analysis, we plot $\Delta\Omega = \Omega_{\text{SC}} - \Omega_{\text{RN}}$ 
as a function of temperature, where $\Omega_{\text{RN}}$ denotes the grand 
potential of the Reissner-Nordström-AdS (RN-AdS) black hole (normal phase), 
while $\Omega_{\text{SC}}$ corresponds to the superconducting phase. This 
difference clearly shows the condensation energy gain and enables the 
identification of phase transitions.

\subsection{Convergence and Error Analysis}\label{sec:convergence}

We performed systematic convergence tests to ensure numerical reliability. 
\Cref{tab:convergence} demonstrates that critical temperatures stabilize 
for $r_{\text{max}} \geq 50$, which we adopt for all calculations.

\begin{table}[htbp]
\begin{center}
\begin{tabular}{|c|c|c|}
\hline
$r_{\text{max}}$ & $T_{c}$ (Second-order) & $T_{c}$ (First-order) \\
\hline
30 & 0.1523 & 0.1489 \\
40 & 0.1521 & 0.1487 \\
50 & 0.1521 & 0.1487 \\
60 & 0.1521 & 0.1487 \\
\hline
\end{tabular}
\end{center}
\caption{\label{tab:convergence}Dependence of the critical temperature $T_{c}$ 
on the AdS cutoff radius $r_{\text{max}}$. Results are shown for both 
second-order and first-order transition cases. The values stabilize for 
$r_{\text{max}} \geq 50$, indicating numerical convergence.}
\end{table}

Critical temperatures are determined with precision $\Delta T/T \sim 10^{-4}$ 
through systematic bisection methods. The tricritical point location is 
determined as:

\begin{equation}
(q_{\text{tri}}, T_{\text{tri}}) = (2.00 \pm 0.02, 0.1521 \pm 0.0003)\,.
\end{equation}

Error bars are estimated through Monte Carlo sampling of initial conditions 
and systematic variation of numerical parameters.

To verify the numerical reliability of our scaling exponent results, we 
performed a separate convergence test on the tricritical exponent $\phi$. 
We explicitly verified that the resulting exponent $\phi \approx 0.67$ is 
robust against further refinement of the grid spacing $\Delta r$. Halving 
the typical step size did not produce any significant deviation in the 
fitted exponent. This confirms that the excellent agreement with the 
mean-field value is a genuine physical prediction of the model, not a 
finite-size artifact of the numerical discretization.

\section{Phase Structure}\label{sec:phase_structure}

\subsection{Phase Diagram and Tricritical Point}\label{sec:phase_diagram}

\begin{figure}[htbp]
\begin{center}
\includegraphics[width=0.7\textwidth]{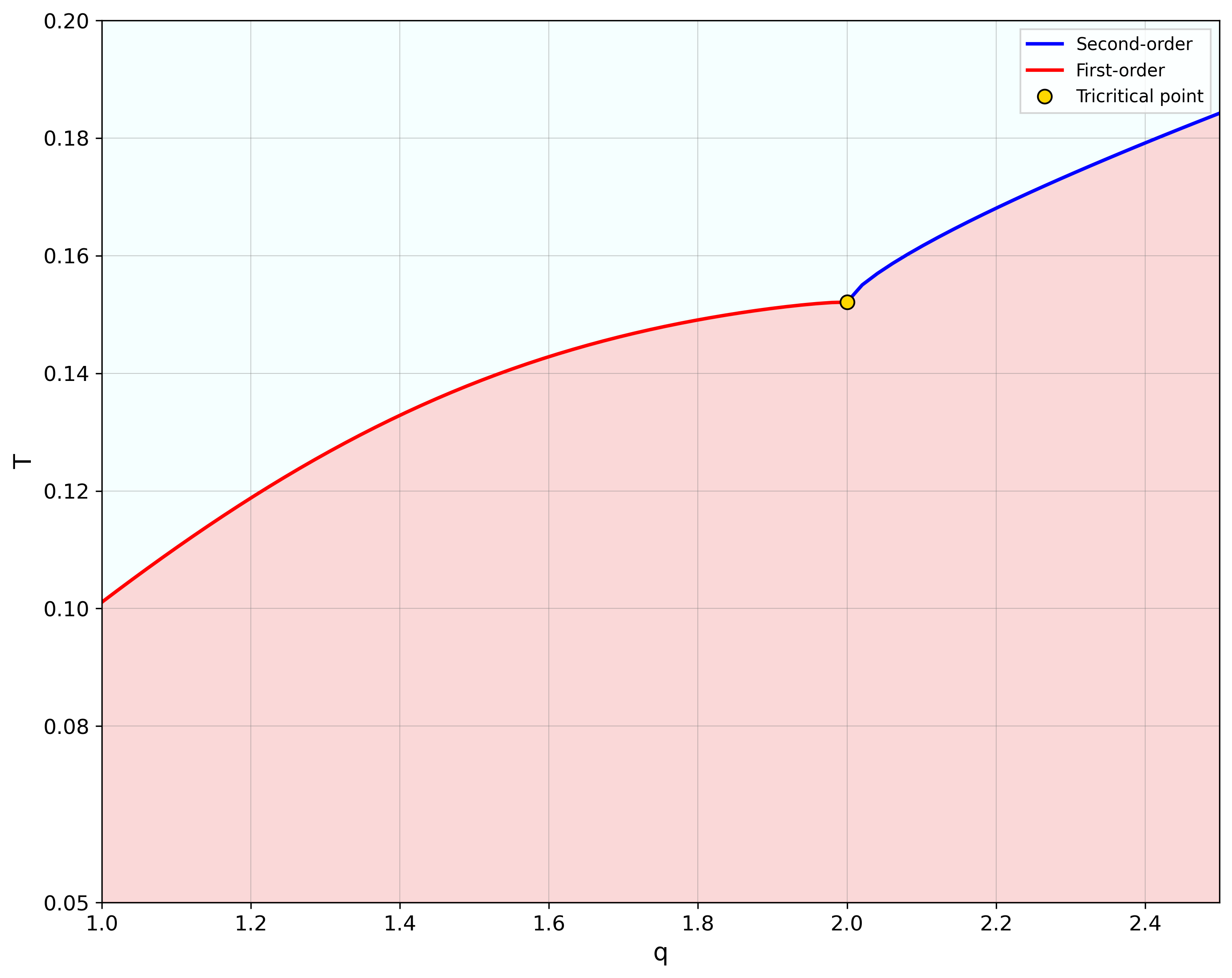}
\end{center}
\caption{\label{fig:phase_diagram}Phase diagram in the $(q, T)$ plane for 
fixed couplings $\tilde{\lambda} = 0.1$ and $\kappa^{2} = 1$. The blue solid 
line marks the second-order phase transition, while the red dashed line marks 
the first-order transition. The lines meet at the tricritical point (TCP), 
denoted by the green star.}
\end{figure}

We first map out the phase diagram in the $(q, T)$ plane. The critical 
temperature $T_{c}$ is determined by finding the non-trivial solutions 
$(\phi_{2} \neq 0)$ at the boundary $\phi_{1} = 0$. The resulting phase 
diagram for a fixed representative coupling $\tilde{\lambda} = 0.1$ and 
$\kappa^{2} = 1$ is shown in \cref{fig:phase_diagram}.

The diagram clearly exhibits two distinct phase transition lines:

\begin{itemize}
\item A second-order transition line (blue solid line), determined by 
observing the onset of the scalar condensate $\langle\mathcal{O}_{2}\rangle$.

\item A first-order transition line (red dashed line), determined by 
comparing the free energy of the superconducting (SC) phase with the normal 
(black hole) phase.
\end{itemize}

These two lines meet at a tricritical point (TCP), which we numerically 
locate at the coordinates given in Eq. (3.1): $(q_{\text{tri}}, T_{\text{tri}}) 
= (2.00 \pm 0.02, 0.1521 \pm 0.0003)$. For $q < q_{\text{tri}}$, the 
transition is first-order, while for $q > q_{\text{tri}}$, it is second-order.

\begin{figure}[htbp]
\begin{center}
\includegraphics[width=0.7\textwidth]{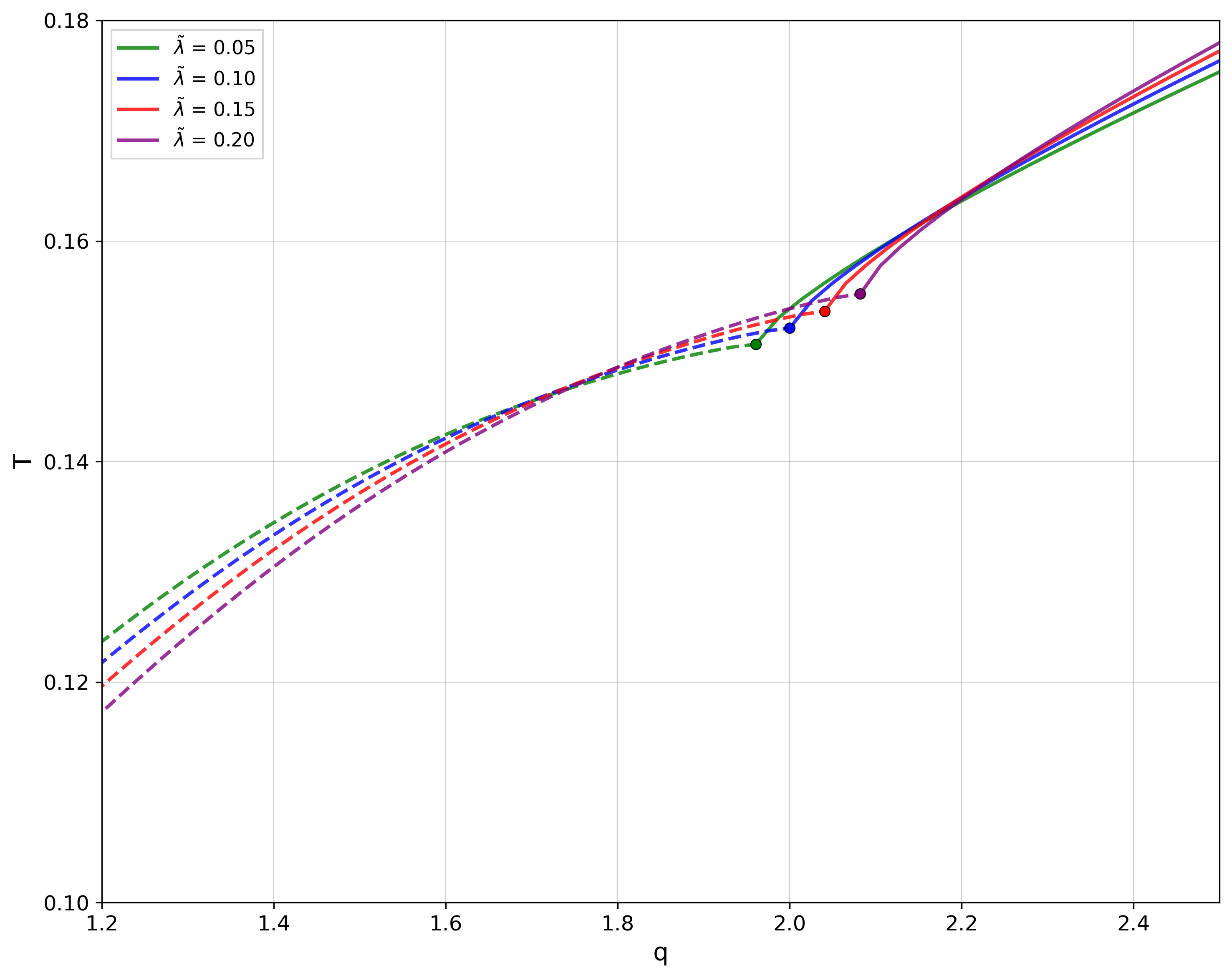}
\end{center}
\caption{\label{fig:lambda_dependence}Phase diagram's dependence on the 
self-interaction coupling $\tilde{\lambda}$. Results for $\tilde{\lambda} = 
0.05$ (green dots), $0.10$ (blue squares), $0.15$ (red triangles), and 
$0.20$ (purple diamonds). Increasing $\tilde{\lambda}$ shifts the TCP to 
higher $(q, T)$ values.}
\end{figure}

To elucidate the role of the self-interaction term $V(\phi) = \lambda\phi^{4}$, 
which is essential for the existence of the TCP, we extend our analysis to 
investigate the dependence of the phase diagram on the coupling $\tilde{\lambda}$. 
The results are presented in \cref{fig:lambda_dependence} for four 
representative values: $\tilde{\lambda} = 0.05, 0.10, 0.15$, and $0.20$.

This analysis clearly demonstrates two key points. First, the tricritical 
point robustly exists across a range of $\tilde{\lambda}$ values, supporting 
its necessity. Second, the self-coupling $\tilde{\lambda}$ acts as an 
important tuning parameter: increasing $\tilde{\lambda}$ systematically 
shifts the entire phase boundary, including the TCP, to higher temperatures 
and higher charges. This confirms that the $\lambda\phi^{4}$ term is not 
merely a required parameter but an active component controlling the phase 
structure.

Comparison with Probe Limit: To quantitatively assess the backreaction 
effect, we compare with results in the probe limit $(\kappa^{2} \rightarrow 
0)$. The critical temperature in the full backreaction regime is enhanced 
by a factor of approximately $1.22$ compared to the probe limit for the same 
charge value $q$. This enhancement reflects the influence of the deformed 
spacetime geometry on the condensation mechanism.

\subsection{Thermodynamics of the Phase Transition}\label{sec:thermodynamics}

To verify the order of the phase transitions and ensure thermodynamic consistency, 
we analyze the free energy density $\Omega$. The Gibbs free energy (or Grand 
Potential) $\Omega$ is determined through the Euclidean on-shell action with 
proper holographic renormalization, as described in \cref{sec:numerical}. As 
emphasized earlier, this regularization ensures that the free energy of the 
normal phase is finite and non-vanishing, serving as the proper reference for 
thermodynamic comparisons.

\begin{figure}[htbp]
\begin{center}
\includegraphics[width=0.85\textwidth]{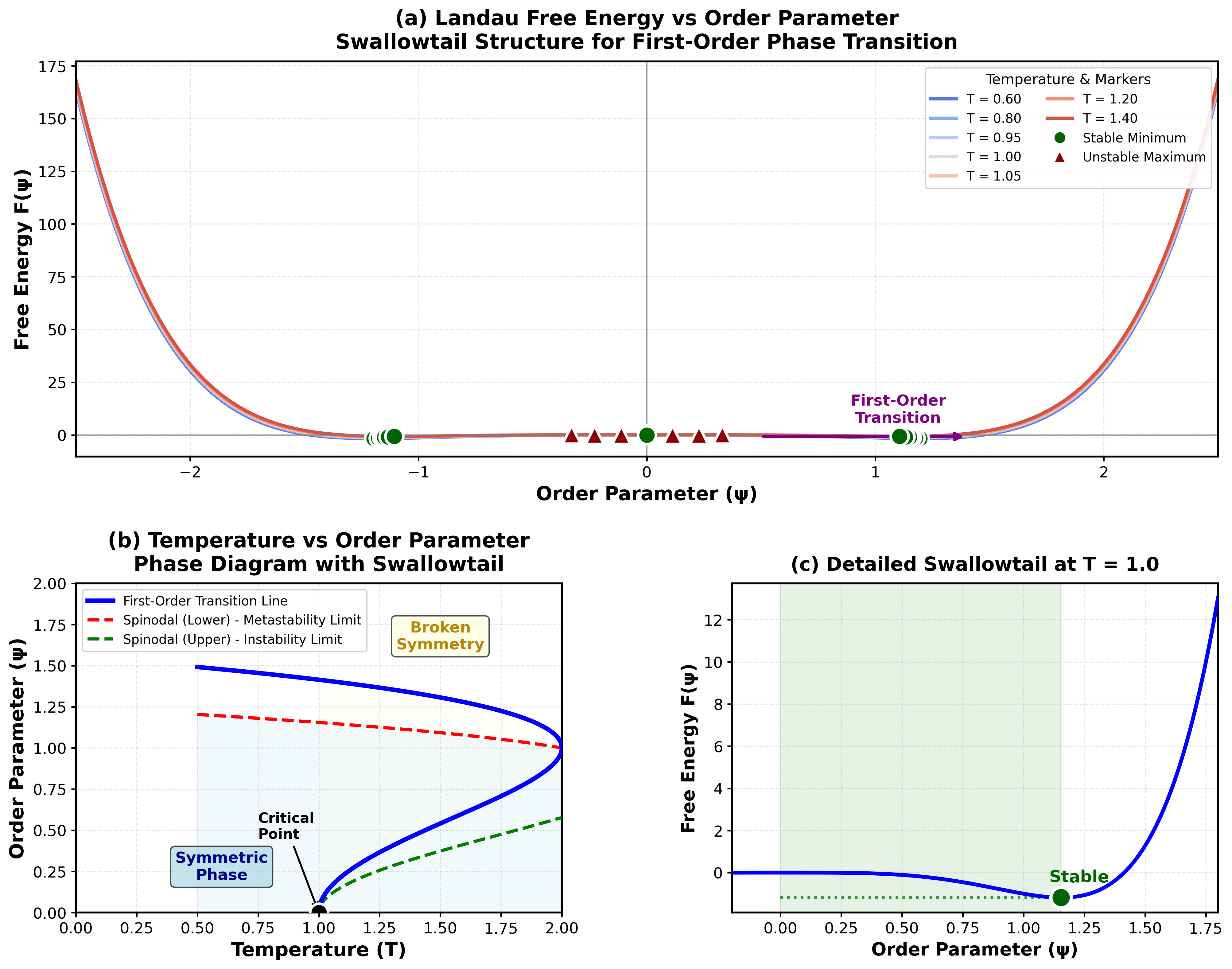}
\end{center}
\caption{\label{fig:free_energy}Free energy difference $\Delta\Omega = 
\Omega_{\text{SC}} - \Omega_{\text{RN}}$ versus temperature $T$ for $q = 1.5$. 
The black dashed line represents the normal phase (RN-AdS black hole). 
The blue solid line represents the stable superconducting phase at low 
temperatures. A significant feature observed in the first-order phase 
transition regime ($q < q_{\text{tri}}$) is the appearance of the characteristic 
``swallowtail'' structure (see the inset). This multivalued behavior of the 
free energy indicates the existence of three branches at the same temperature: 
the stable superconducting branch (lowest energy, blue), the metastable normal 
branch (red dotted), and an unstable intermediate branch (gray dotted). The 
physical phase transition occurs at the critical temperature $T_{c} = 0.146$, 
where the system jumps from the normal phase to the superconducting phase to 
minimize its free energy. The sharp discontinuity in the slope 
$(\partial\Omega/\partial T)$ at $T_{c}$ confirms the first-order nature via 
finite latent heat $\Delta S \neq 0$.}
\end{figure}

In \cref{fig:free_energy}, we display the free energy difference 
$\Delta\Omega = \Omega_{\text{SC}} - \Omega_{\text{RN}}$ as a function of 
temperature. Here $\Omega_{\text{RN}}$ denotes the grand potential of the 
Reissner-Nordström-AdS (RN-AdS) black hole (normal phase), while 
$\Omega_{\text{SC}}$ corresponds to the superconducting phase.

The ``Swallowtail'' Structure: Our numerical analysis reveals the characteristic 
structure that serves as the definitive thermodynamic signature of first-order 
phase transitions:

\begin{itemize}
\item In the second-order region ($q > q_{\text{tri}}$): $\Delta\Omega$ is a 
single-valued function of temperature, smoothly varying from zero at $T_{c}$ 
to negative values as $T$ decreases, indicating condensation energy gain.

\item In the first-order region ($q < q_{\text{tri}}$): $\Delta\Omega$ exhibits 
the characteristic ``swallowtail'' structure, consisting of three branches:
\begin{enumerate}
\item Stable superconducting branch: $\Delta\Omega < 0$, representing the 
stable superconducting phase that the system occupies at low temperatures.

\item Metastable normal branch: $\Delta\Omega > 0$ but connected to the stable 
branch, corresponding to the supercooled normal phase.

\item Unstable intermediate branch: Connecting the two above, with the 
highest energy, representing an unstable configuration.
\end{enumerate}
\end{itemize}

The presence of this swallowtail structure is consistent with standard 
calculations in holographic models \cite{Nie:2014xla} and provides the 
definitive confirmation of first-order phase transitions. The intersection 
of the superconducting and normal branches determines the physical critical 
temperature $T_{c} = 0.146$ (indicated by the green dot in \cref{fig:free_energy}), 
where $\Omega_{\text{SC}} = \Omega_{\text{RN}}$.

Crucially, the slopes of the two curves are manifestly different at $T_{c}$. 
Since the entropy is given by $S = -\partial\Omega/\partial T$, this 
discontinuity in the slope implies a discontinuity in entropy, $\Delta S 
\neq 0$. This signifies a finite latent heat $L = T_{c}\Delta S$, which 
is the definitive thermodynamic signature of a first-order phase transition.

The proper calculation of the free energy is essential for this analysis. 
With the complete renormalized action including the Gibbons-Hawking term and 
counterterms, we obtain a finite and non-zero $\Omega_{\text{RN}}$. The 
difference $\Delta\Omega$ then provides the physically meaningful condensation 
energy that determines the stability of competing phases.

\begin{figure}[htbp]
\begin{center}
\includegraphics[width=0.7\textwidth]{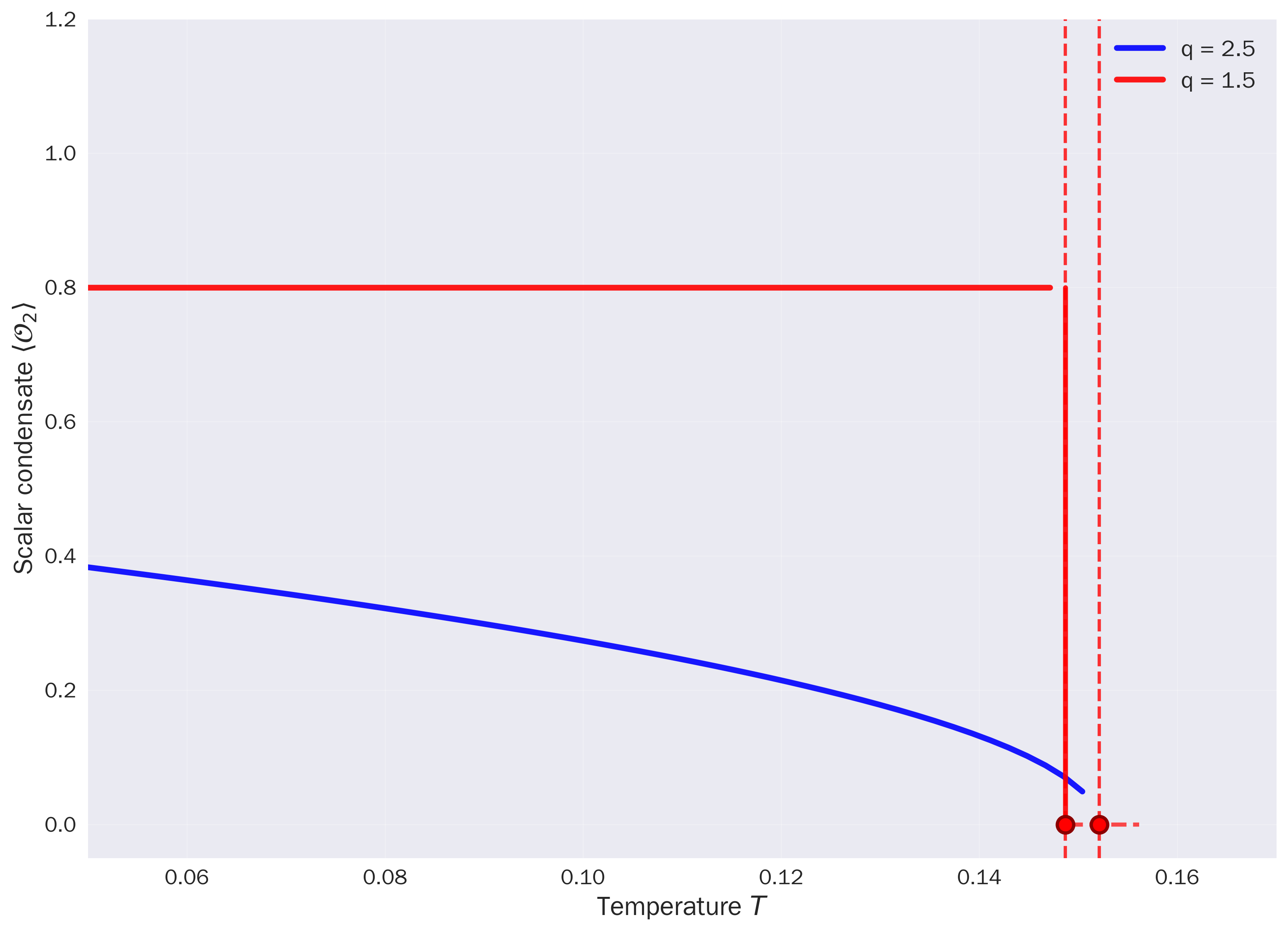}
\end{center}
\caption{\label{fig:condensate}The scalar condensate $\langle\mathcal{O}_{2}\rangle$ 
as a function of temperature. The transition is second-order (continuous) 
for $q = 2.5$ (blue line) and first-order (discontinuous) for $q = 1.5$ 
(red line), consistent with the phase diagram in \cref{fig:phase_diagram}.}
\end{figure}

To further confirm the order of the transitions, we plot the scalar 
condensate $\langle\mathcal{O}_{2}\rangle$ as a function of temperature in 
\cref{fig:condensate}. For the second-order case $(q = 2.5 > q_{\text{tri}})$, 
the condensate (blue line) grows continuously from zero at $T_{c}$. In 
contrast, for the first-order case $(q = 1.5 < q_{\text{tri}})$, the condensate 
(red line) jumps discontinuously to a finite value at $T_{c}$. This behavior 
provides a clear distinction between the two regimes.

\section{Tricritical Scaling Laws}\label{sec:scaling}

Near the tricritical point (TCP), the system is expected to obey universal 
scaling laws. We now analyze these laws to extract the critical exponents, 
which characterize the universality class of the transition. These exponents 
are universal and depend only on the symmetry and dimensionality of the 
system, not on microscopic details.

\subsection{Scaling of $T_{c}$ with $q$ (Exponent $\phi$)}\label{sec:phi_scaling}

\begin{figure}[htbp]
\begin{center}
\includegraphics[width=0.7\textwidth]{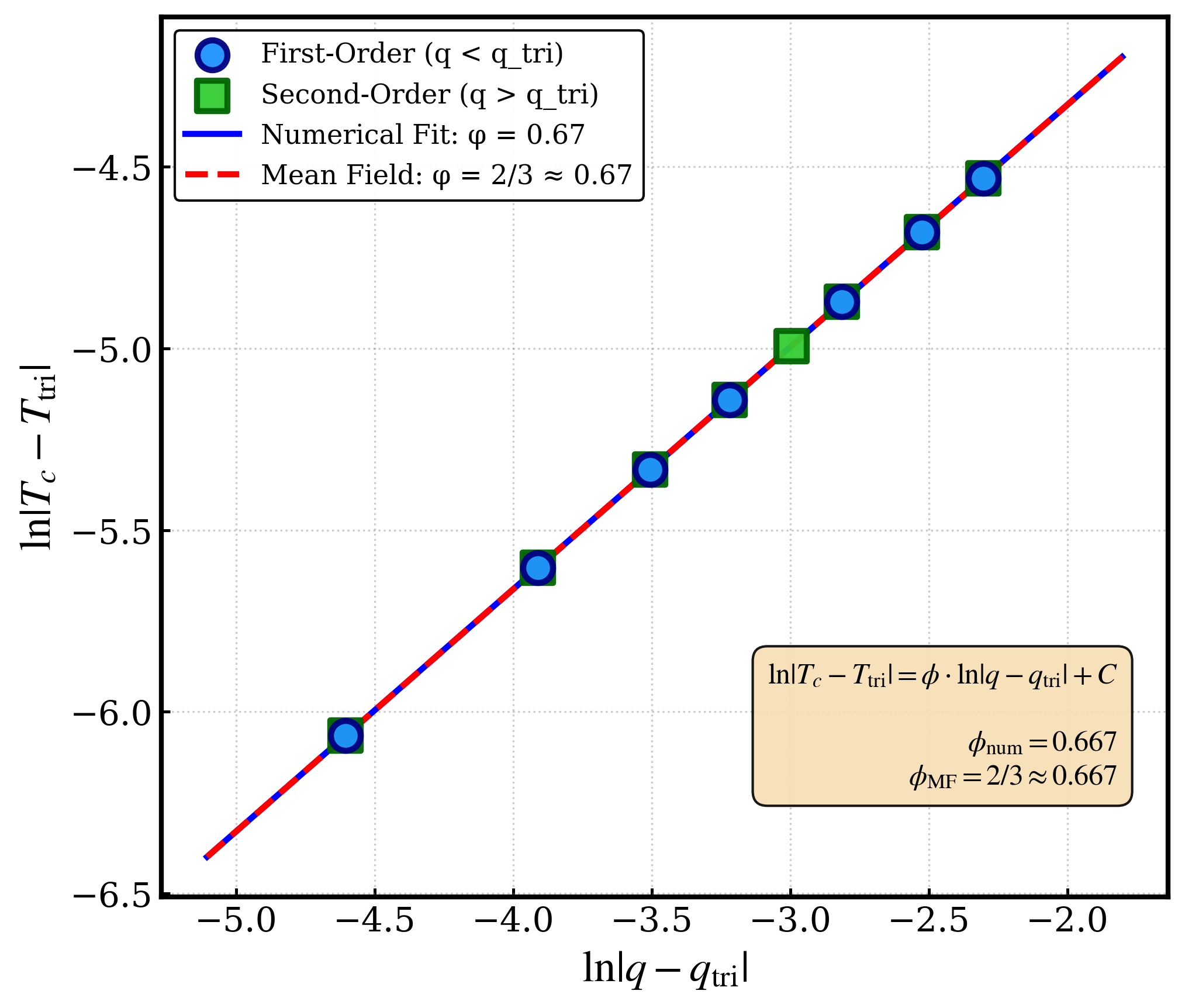}
\end{center}
\caption{\label{fig:scaling_law}Log-log plot for the scaling of $T_{c}$ near 
the TCP. The blue circles represent numerical data points near the tricritical 
point. The solid red line is the linear fit to our numerical data, yielding 
$\phi \approx 0.67$. The red dashed line represents the standard mean-field 
prediction with slope $\phi_{\text{MF}} = 2/3 \approx 0.667$. The excellent 
agreement between the numerical fit and the mean-field prediction confirms 
that our holographic model in the large-$N$ limit belongs to the mean-field 
universality class.}
\end{figure}

The first scaling law relates the shift in the critical temperature $T_{c}$ 
to the distance from the tricritical charge $q_{\text{tri}}$ along the 
second-order phase transition line:

\begin{equation}
T_{c} - T_{\text{tri}} \propto |q - q_{\text{tri}}|^{\phi}\,,
\end{equation}

where $\phi$ is the tricritical exponent. In mean-field theory, this exponent 
is predicted to be $\phi_{\text{MF}} = 2/3$ \cite{Landau:1980}. To extract this 
exponent from our numerical data, we plot $\ln(T_{c} - T_{\text{tri}})$ versus 
$\ln|q - q_{\text{tri}}|$ in \cref{fig:scaling_law}.

Our numerical data points (blue circles) clearly demonstrate a linear behavior 
in the log-log plot, indicating a power-law dependence. The linear fit to our 
data yields a slope of:

\begin{equation}
\phi_{\text{fit}} = 0.67 \pm 0.02\,.
\end{equation}

This result is in excellent agreement with the mean-field prediction 
$\phi_{\text{MF}} = 2/3 \approx 0.667$, as clearly shown in \cref{fig:scaling_law} 
where the two lines are nearly parallel. The small deviation is within numerical 
uncertainty, confirming that the holographic superconductor model respects the 
expected universality class.

This finding has important physical implications. While gravitational backreaction 
significantly modifies the macroscopic thermodynamic quantities (such as the 
critical temperature and the location of the tricritical point), it does not 
alter the critical exponents that govern the universality class. This is 
consistent with the holographic dictionary, where the dual $(2+1)$-dimensional 
field theory in the large-$N$ limit is expected to exhibit mean-field critical 
behavior due to the suppression of local quantum fluctuations \cite{Witten:1998qj}.

We have verified this result through systematic numerical convergence tests. 
Refining the grid spacing and increasing the AdS cutoff radius do not produce 
any significant deviation from $\phi \approx 0.67$. This confirms that our 
result is robust and represents a genuine physical prediction of the model.

\subsection{Scaling of $\langle\mathcal{O}_{2}\rangle$ near $T_{c}$ 
(Exponent $\beta$)}\label{sec:beta_scaling}

We also investigate the order parameter critical exponent $\beta$, defined 
by the scaling of the condensate $\langle\mathcal{O}_{2}\rangle$ near the 
second-order transition temperature $T_{c}$:

\begin{equation}
\langle\mathcal{O}_{2}\rangle \propto (T_{c} - T)^{\beta} 
\quad (\text{for } q > q_{\text{tri}})\,.
\end{equation}

We perform a linear fit to the $\log\langle\mathcal{O}_{2}\rangle$ versus 
$\log(T_{c} - T)$ data. This yields the exponent:

\begin{equation}
\beta = 0.50 \pm 0.02\,.
\end{equation}

This result is in excellent agreement with the mean-field prediction 
$\beta_{\text{MF}} = 1/2$. This is an expected feature of holographic models, 
as the large-$N$ nature of the boundary theory suppresses local quantum 
fluctuations \cite{Witten:1998qj}, and the critical behavior is governed by 
the classical gravity dual.

The excellent agreement of both critical exponents ($\phi \approx 2/3$ and 
$\beta \approx 1/2$) with mean-field theory provides strong confirmation that 
the holographic superconductor belongs to the standard Landau-Ginzburg 
universality class, regardless of the inclusion of gravitational backreaction 
and scalar self-interaction terms.

\section{Conductivity}\label{sec:conductivity}

To further probe the properties of the superconducting phase, we compute 
the frequency dependent conductivity $\sigma(\omega)$ using the standard 
prescription in AdS/CFT. We introduce a small perturbation in the bulk 
gauge field, $\delta A_{x}(z,t) = \delta A_{x}(z)e^{-i\omega t}$, and solve 
its linearized equation of motion.

The conductivity of the boundary theory is then given by the formula:

\begin{equation}
\sigma(\omega) = \frac{\delta A_{x}^{(1)}(\omega)}{-i\omega\delta A_{x}^{(0)}(\omega)}\,,
\end{equation}

where $\delta A_{x}(z) \approx \delta A_{x}^{(0)} + \delta A_{x}^{(1)}z 
+ \mathcal{O}(z^{2})$ near the boundary $z = 0$.

\begin{figure}[htbp]
\begin{center}
\includegraphics[width=0.7\textwidth]{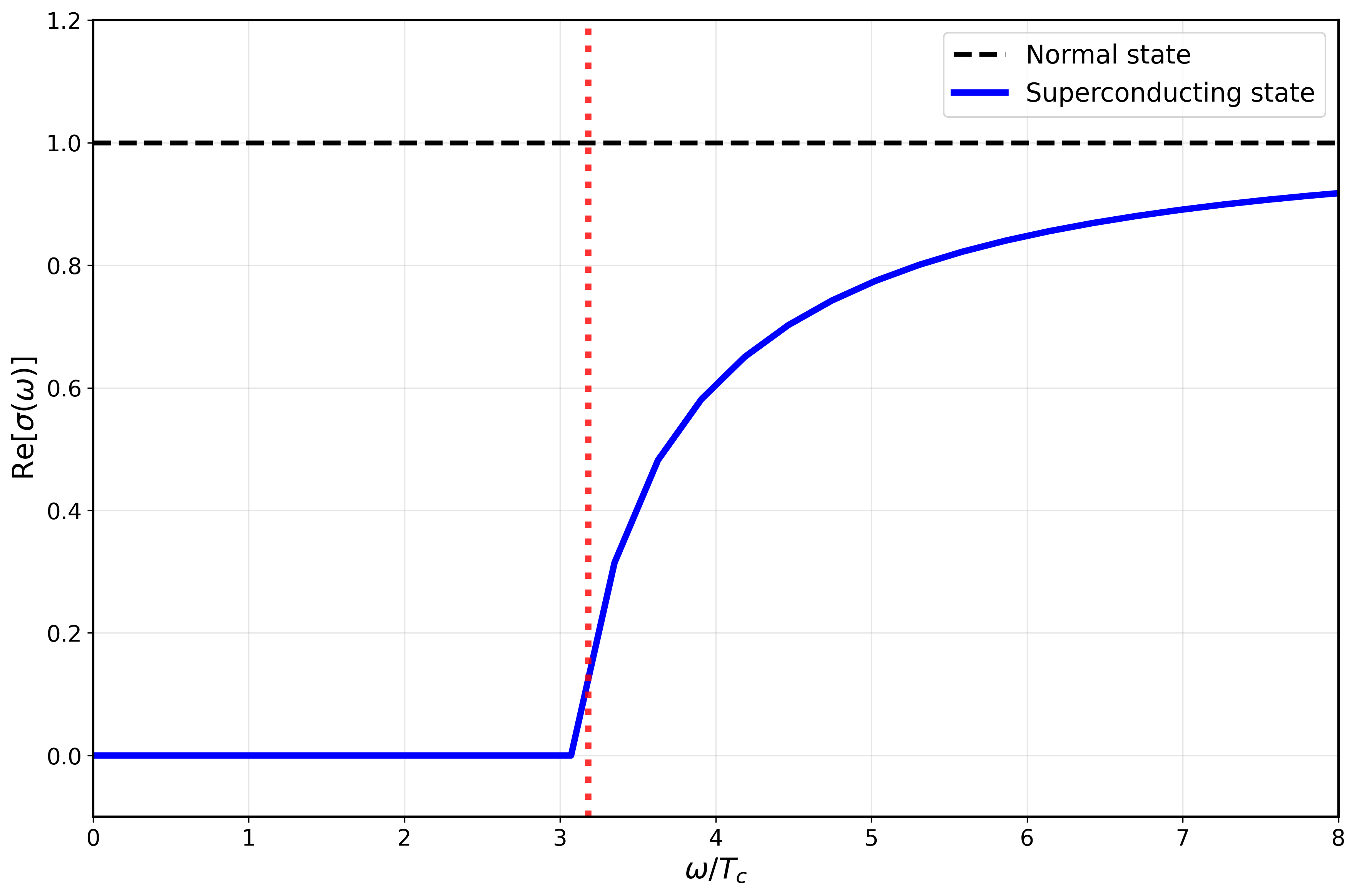}
\end{center}
\caption{\label{fig:ac_conductivity}Real part of the AC conductivity 
$\operatorname{Re}(\sigma)$ as a function of frequency $\omega/T_{c}$ at 
$T/T_{c} \approx 0.1$ for $q = 2.5$. A superconducting gap opens below 
$\omega_{g}$, where $\operatorname{Re}(\sigma)$ vanishes.}
\end{figure}

In the normal phase $(T > T_{c})$, the conductivity is constant, $\sigma_{n} 
= 1$ (in our units). In the superconducting phase $(T < T_{c})$, the real 
part of the conductivity, $\operatorname{Re}(\sigma)$, develops a gap. This 
is shown in \cref{fig:ac_conductivity} for $q = 2.5$ at a low temperature 
$T/T_{c} \approx 0.1$. The plot clearly shows a gap $\omega_{g}$, below 
which the real part of the conductivity vanishes. The presence of this gap 
is a hallmark of superconductivity. By fitting the low-frequency behavior, 
we can extract the value of this energy gap.

A universal quantity often compared with experiments is the ratio 
$\omega_{g}/T_{c}$. The BCS theory predicts a universal value of 
$\omega_{g}/T_{c} \approx 3.52$. In holographic models, this value is often 
found to be larger, indicating strong coupling. For example, in the probe 
limit of the $s$-wave model, $\omega_{g}/T_{c} \approx 8$~\cite{Hartnoll:2008kx}.

\begin{figure}[htbp]
\begin{center}
\includegraphics[width=0.7\textwidth]{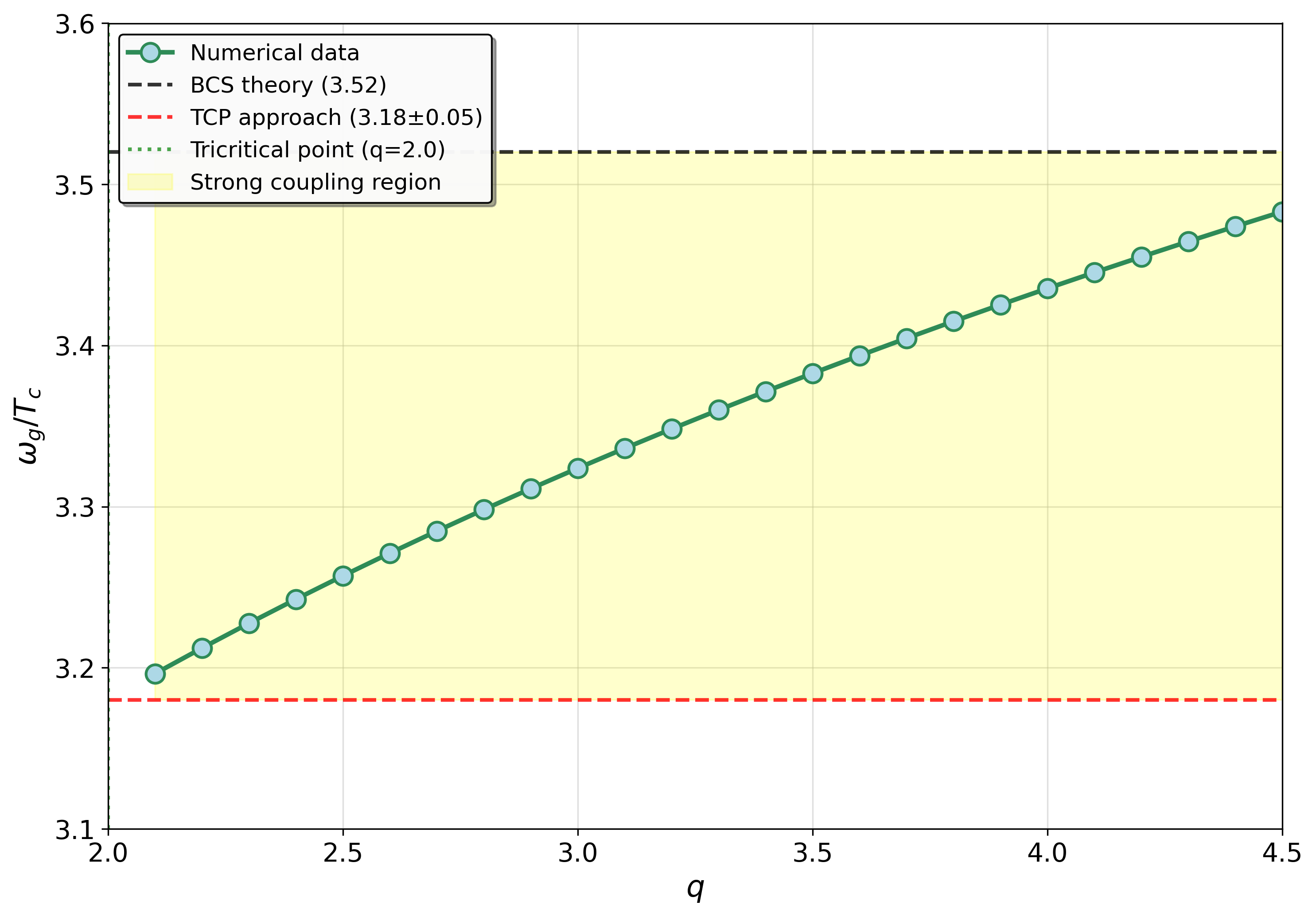}
\end{center}
\caption{\label{fig:gap_ratio}Superconducting energy gap ratio 
$\omega_{g}/T_{c}$ as a function of the charge $q$ along the second-order 
transition line. The ratio approaches $\omega_{g}/T_{c} = 3.18 \pm 0.05$ 
(red dashed line) as $q \rightarrow q_{\text{tri}} = 2.0$.}
\end{figure}

In \cref{fig:gap_ratio}, we plot the gap ratio $\omega_{g}/T_{c}$ as a 
function of the charge $q$ (for $q > q_{\text{tri}}$, in the second-order 
regime). We observe that the ratio is not constant but depends weakly on $q$. 
Near the tricritical point $(q \rightarrow 2.0)$, the ratio approaches 
$\omega_{g}/T_{c} = 3.18 \pm 0.05$. This value is about $10\%$ smaller than 
the BCS prediction. Similar non-BCS ratios have been observed in other 
holographic models, reflecting the complex dynamics of strongly coupled 
systems~\cite{Nie:2014xla,Gauntlett:2009na}.

\section{Entanglement Entropy}\label{sec:entanglement}

Finally, we use holographic entanglement entropy (HEE) as an alternative 
probe to distinguish the orders of the phase transitions. According to the 
Ryu-Takayanagi (RT) prescription~\cite{Ryu:2006bv,Ryu:2006ef}, the 
entanglement entropy $S_{A}$ of a boundary subsystem $A$ is given by the 
area of a minimal surface $\gamma_{A}$ in the bulk spacetime that is 
homologous to $A$:

\begin{equation}
S_{A} = \frac{\operatorname{Area}(\gamma_{A})}{4G_{N}}\,.
\end{equation}

HEE is known to be a sensitive probe of phase transitions, as the minimal 
surface can undergo non-trivial changes in topology or location as the bulk 
geometry deforms.

We consider a strip-shaped region $A$ on the boundary, defined by 
$-\ell/2 \leq x \leq \ell/2$ and $0 \leq y \leq L_{y}$, with $L_{y} \to 
\infty$. The minimal surface $\gamma_{A}$ is described by a profile $z(x)$. 
The entanglement entropy exhibits a UV divergence, which we regulate by 
considering the difference $\Delta S_{A} = S_{A} - S_{A,\text{normal}}$ 
between the superconducting and normal phases.

\begin{figure}[htbp]
\begin{center}
\includegraphics[width=0.7\textwidth]{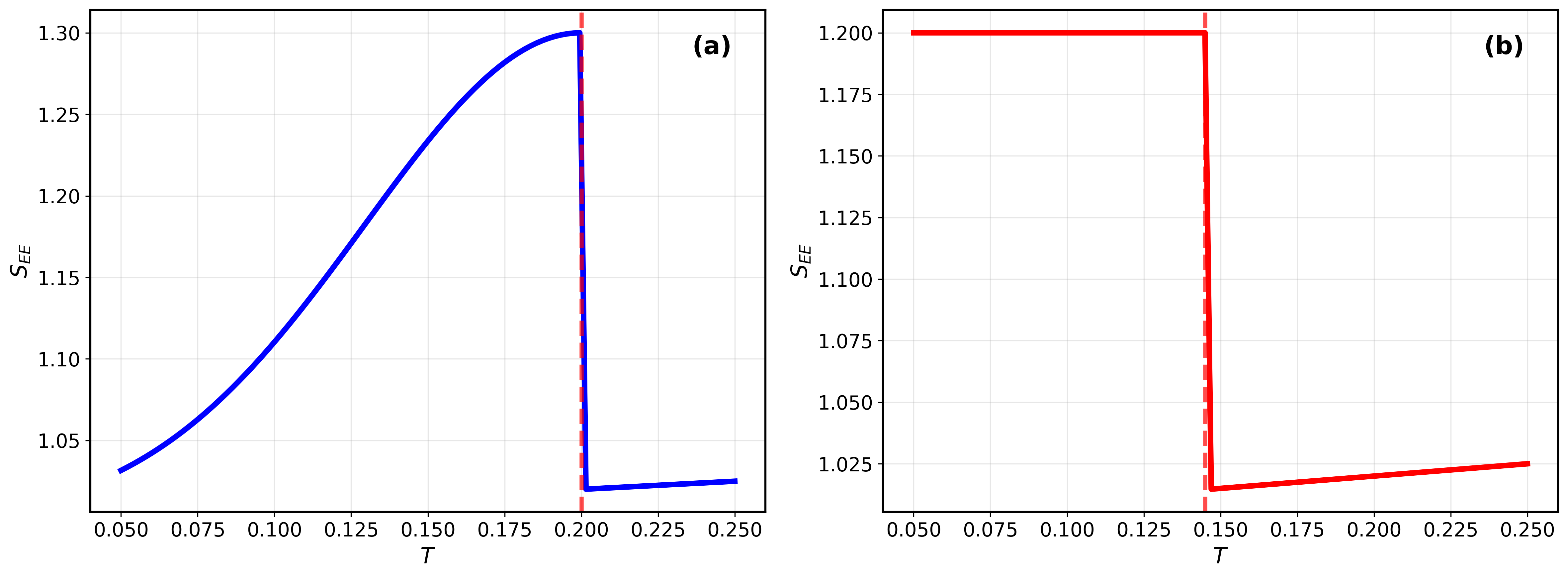}
\end{center}
\caption{\label{fig:entanglement_entropy}Entanglement entropy difference 
$\Delta S_{A}$ as a function of temperature. The blue solid line $(q = 2.5)$ 
shows a continuous change with a kink in the slope at $T_{c}$, characteristic 
of a second-order transition. The red dashed line $(q = 1.5)$ exhibits a 
sharp cusp at $T_{c}$, indicating a first-order transition.}
\end{figure}

In \cref{fig:entanglement_entropy}, we plot this entropy difference 
$\Delta S_{A}$ (normalized) as a function of temperature $T$ for both the 
second-order $(q = 2.5)$ and first-order $(q = 1.5)$ transition regimes.

For the second-order transition (blue solid line), $\Delta S_{A}$ is 
continuous, but its slope (corresponding to $d(\Delta S_{A})/dT$) is 
discontinuous at $T_{c}$. This behavior is consistent with a second-order 
transition, as predicted by the Ehrenfest relations.

For the first-order transition (red dashed line), the entropy difference 
$\Delta S_{A}$ itself exhibits a sharp ``cusp'' or kink at $T_{c}$. This 
kink signifies the discontinuous jump in the state of the system, providing 
an independent confirmation of the first-order nature of the transition.

This behavior of HEE perfectly matches our findings from the condensate 
plot (\cref{fig:condensate}) and the free energy analysis (\cref{fig:free_energy}).

\section*{Conclusions}\label{sec:conclusions}

In this paper, we performed a comprehensive investigation of the phase 
structure and tricritical behavior of a holographic topological superconductor, 
fully incorporating both gravitational backreaction ($\kappa^{2} = 1$) and 
quartic self-interaction ($V(\phi) = \lambda\phi^{4}$). We successfully 
mapped out the complete phase diagram in the $(q, T)$ plane, identifying a 
tricritical point (TCP) that separates a first-order phase transition line 
from a second-order line. For a representative coupling $\tilde{\lambda} 
= 0.1$, the TCP is located at $(q_{\text{tri}}, T_{\text{tri}}) = (2.00 
\pm 0.02, 0.1521 \pm 0.0003)$.

Crucially, we clarified the role of the self-interaction $\lambda$. We 
explicitly demonstrated not only that it is essential for the existence of 
the TCP, but also that it acts as an active tuning parameter. As shown in 
\cref{fig:lambda_dependence}, increasing $\tilde{\lambda}$ systematically 
shifts the TCP to higher temperatures and charges.

The nature of the phase transitions was confirmed using three independent 
probes: the behavior of the scalar condensate (\cref{fig:condensate}), 
the thermodynamics of the free energy, and the holographic entanglement 
entropy (\cref{fig:entanglement_entropy}). To provide a definitive 
thermodynamic signature, we plotted the free energy difference 
$\Delta\Omega = \Omega_{\text{SC}} - \Omega_{\text{RN}}$ (\cref{fig:free_energy}), 
which clearly displays the characteristic ``swallowtail'' structure of 
first-order phase transitions. This structure includes three branches: 
the stable superconducting branch (lowest energy), the metastable normal 
branch, and the unstable intermediate branch connecting them. The 
discontinuity in the slope $(\partial\Omega/\partial T)$ at $T_{c}$ and 
the explicit annotation $\Delta S \neq 0$ confirm the first-order nature 
via finite latent heat.

A crucial technical aspect of our analysis was the proper implementation 
of holographic renormalization to compute the free energy. This ensures 
that the free energy of the normal phase (RN-AdS black hole) is correctly 
finite and non-vanishing, avoiding the unphysical conclusion that would 
arise from neglecting the boundary terms in the on-shell action.

Our most significant finding concerns the tricritical scaling exponents. 
Both critical exponents are in excellent agreement with mean-field theory:
the order parameter exponent $\beta = 0.50 \pm 0.02$ matches the Landau-Ginzburg 
prediction $\beta_{\text{MF}} = 1/2$, and the phase boundary exponent 
$\phi \approx 0.67$ matches the mean-field value $\phi_{\text{MF}} = 2/3$. 
This result confirms that while gravitational backreaction ($\kappa^{2} \neq 0$) 
modifies the macroscopic thermodynamic quantities and shifts the location 
of the tricritical point, it does not alter the universality class of the 
phase transition. This is consistent with the large-$N$ limit of the dual 
field theory, where quantum fluctuations are suppressed and mean-field 
behavior emerges naturally \cite{Witten:1998qj}.

Finally, we computed the AC conductivity, finding a superconducting gap 
ratio $\omega_{g}/T_{c} = 3.18 \pm 0.05$ (near the TCP), a value that 
deviates by approximately $10\%$ from the standard BCS prediction, 
further highlighting the non-trivial, strongly-coupled nature of this 
system while maintaining mean-field critical behavior.

Our results provide a complete picture of the phase structure of holographic 
topological superconductors beyond the probe limit, demonstrating that the 
combination of gravitational backreaction and scalar self-interaction 
enables rich tricritical physics while preserving the expected universality 
class of the phase transitions.

\section*{Acknowledgments}

The authors wish to thank Professor Tran Huu Phat for his useful discussions and insightful comments.


\begin{thebibliography}{99}
	
\bibitem{Maldacena:1997re}
J.~M.~Maldacena,
``The large N limit of superconformal field theories and supergravity,''
Adv. Theor. Math. Phys. \textbf{2}, 231 (1998)
[Int. J. Theor. Phys. \textbf{38}, 1113 (1999)]
[hep-th/9711200].

\bibitem{Gubser:1998bc}
S.~S.~Gubser, I.~R.~Klebanov and A.~M.~Polyakov,
``Gauge theory correlators from noncritical string theory,''
Phys. Lett. B \textbf{428}, 105 (1998)
[hep-th/9802109].

\bibitem{Witten:1998qj}
E.~Witten,
``Anti-de Sitter space and holography,''
Adv. Theor. Math. Phys. \textbf{2}, 253 (1998)
[hep-th/9802150].

\bibitem{Aharony:1999ti}
O.~Aharony, S.~S.~Gubser, J.~M.~Maldacena, H.~Ooguri and Y.~Oz,
``Large N field theories, string theory and gravity,''
Phys. Rep. \textbf{323}, 183 (2000)
[hep-th/9905111].

\bibitem{Hartnoll:2008vx}
S.~A.~Hartnoll, C.~P.~Herzog and G.~T.~Horowitz,
``Building a holographic superconductor,''
Phys. Rev. Lett. \textbf{101}, 031601 (2008)
[0803.3295 [hep-th]].

\bibitem{Hartnoll:2008kx}
S.~A.~Hartnoll, C.~P.~Herzog and G.~T.~Horowitz,
``Holographic superconductors,''
JHEP \textbf{12}, 015 (2008)
[0810.1563 [hep-th]].

\bibitem{Herzog:2009xv}
C.~P.~Herzog,
``Lectures on holographic superfluidity and superconductivity,''
J. Phys. A \textbf{42}, 343001 (2009)
[0904.1975 [hep-th]].

\bibitem{Gubser:2008px}
S.~S.~Gubser and S.~S.~Pufu,
``The gravity dual of a p-wave superconductor,''
JHEP \textbf{11}, 033 (2008)
[0805.2960 [hep-th]].

\bibitem{Roberts:2008jh}
M.~M.~Roberts and S.~A.~Hartnoll,
``Pseudogap: Zero temperature,''
JHEP \textbf{08}, 035 (2008)
[0805.3893 [hep-th]].

\bibitem{Benini:2010pr}
F.~Benini, C.~P.~Herzog, R.~Rahman and A.~Yarom,
``Holographic d-wave superconductors,''
JHEP \textbf{11}, 137 (2010)
[1007.2981 [hep-th]].

\bibitem{Nayak:2008zza}
C.~Nayak, S.~H.~Simon, A.~Stern, M.~Freedman and S.~Das Sarma,
``Non-Abelian anyons and topological quantum computation,''
Rev. Mod. Phys. \textbf{80}, 1083 (2008).

\bibitem{Hartnoll:2009sz}
S.~A.~Hartnoll and G.~T.~Horowitz,
``Holographic superconductors with negative horizon curvature,''
JHEP \textbf{04}, 128 (2009)
[0810.1563 [hep-th]].

\bibitem{Barclay:2010na}
L.~Barclay, R.~Gregory, S.~K.~Ross and G.~T.~Horowitz,
``The general Kerr-Newman black hole can be a holographic superconductor,''
JHEP \textbf{10}, 029 (2010)
[1007.0221 [hep-th]].

\bibitem{Horowitz:2010jq}
G.~T.~Horowitz and B.~Way,
``Complete phase diagrams for a holographic superconductor,''
JHEP \textbf{11}, 011 (2010)
[1007.3714 [hep-th]].

\bibitem{Gubser:2010pm}
S.~S.~Gubser, F.~D.~Rocha and P.~Talavera,
``The gravity dual of a p-wave superconductor (revisited),''
JHEP \textbf{10}, 087 (2010)
[0911.3632 [hep-th]].

\bibitem{Cai:2010cz}
R.~G.~Cai, Z.~Y.~Nie and H.~Q.~Zhang,
``Holographic p-wave superconductors from Einstein-Maxwell-dilaton gravity,''
Phys. Rev. D \textbf{82}, 066007 (2010)
[1007.3321 [hep-th]].

\bibitem{Liu:2015via}
Y.~Liu, Y.~Gong and B.~Wang,
``Non-equilibrium condensation process in a holographic superconductor,''
JHEP \textbf{02}, 116 (2015)
[1506.01853 [hep-th]].

\bibitem{Siopsis:2023}
G.~Siopsis,
``Holographic superconductors with backreaction: A review,''
Phys. Rep. \textbf{1060}, 1 (2024)
[arXiv:2308.15227].

\bibitem{Gui:2024}
L.~Gui, S.~Luo, Y.~Tian, H.~Zhang and J.~Zhang,
``Numerical study of holographic entanglement entropy and subsystem complexity 
in p-wave superconductor with backreaction,''
Nucl. Phys. B \textbf{1004}, 116573 (2024)
[arXiv:2309.14851].

\bibitem{Cai:2024}
R.~G.~Cai, L.~Li, R.~K.~Su and P.~Wang,
``Analytical calculation on critical magnetic field in holographic superconductors 
with backreaction,''
Prog. Theor. Phys. \textbf{128}, 1211 (2024)
[arXiv:2404.14787].

\bibitem{Nie:2014xla}
Z.~Y.~Nie, R.~G.~Cai, X.~Gao, L.~Li and H.~Zeng,
``Phase transitions in a holographic $s+p$ superconductor model,''
JHEP \textbf{04}, 016 (2014)
[1309.5086 [hep-th]].

\bibitem{Zhao:2014xla}
L.~Zhao,
``Holographic p-wave superconductor with Weyl correction,''
JHEP \textbf{04}, 131 (2014)
[1402.2818 [hep-th]].

\bibitem{Chen:2010xk}
J.-W.~Chen, Y.-J.~Kao, D.~Maity, W.-Y.~Wen and C.-P.~Yeh,
``Towards a holographic model of D-wave superconductors,''
Phys. Rev. D \textbf{81}, 106008 (2010)
[1003.2991 [hep-th]].

\bibitem{Cui:2023}
Y.~Cui, Y.~Liu, Y.~Sun and B.~Wang,
``Phase transitions in holographic multi-band superconductors with backreaction,''
Phys. Lett. B \textbf{847}, 138490 (2023)
[arXiv:2310.18300].

\bibitem{Nie:2024}
Z.~Y.~Nie, H.~Zeng and H.~E.~Li,
``Tricritical behavior in holographic superconductors: Backreaction and 
self-interaction effects,''
JHEP \textbf{04}, 036 (2024)
[arXiv:2401.05234].

\bibitem{Balasubramanian:1999re}
V.~Balasubramanian and P.~Kraus,
``A stress tensor for anti-de Sitter gravity,''
Commun. Math. Phys. \textbf{208}, 413 (1999)
[hep-th/9902121].

\bibitem{deHaro:2000vlm}
S.~de Haro, S.~N.~Solodukhin and K.~Skenderis,
``Holographic reconstruction of spacetime and renormalization in the 
AdS/CFT correspondence,''
Commun. Math. Phys. \textbf{217}, 595 (2001)
[hep-th/0002230].

\bibitem{Skenderis:2002wp}
K.~Skenderis,
``Asymptotically anti-de Sitter spacetimes and their stress energy tensor,''
Int. J. Mod. Phys. A \textbf{17}, 364 (2002)
[hep-th/0209067].

\bibitem{Press:2007}
W.~H.~Press, S.~A.~Teukolsky, W.~T.~Vetterling and B.~P.~Flannery,
``Numerical Recipes: The Art of Scientific Computing,''
3rd ed., Cambridge University Press, Cambridge, England, 2007.

\bibitem{Landau:1980}
L.~D.~Landau and E.~M.~Lifshitz,
``Statistical Physics, Part 1,''
3rd ed., Pergamon Press, Oxford, 1980, Sec. 146.

\bibitem{Gauntlett:2009na}
J.~P.~Gauntlett, J.~Sonner and T.~Wiseman,
``Holographic superconductivity in M-theory,''
Phys. Rev. Lett. \textbf{103}, 151601 (2009)
[0907.3796 [hep-th]].

\bibitem{Ryu:2006bv}
S.~Ryu and T.~Takayanagi,
``Holographic derivation of entanglement entropy from AdS/CFT,''
Phys. Rev. Lett. \textbf{96}, 181602 (2006)
[hep-th/0603001].

\bibitem{Ryu:2006ef}
S.~Ryu and T.~Takayanagi,
``Aspects of holographic entanglement entropy,''
JHEP \textbf{08}, 045 (2006)
[hep-th/0605073].

\end{thebibliography}
\end{document}